\documentstyle[twoside,graphicx]{article}

\catcode`\@=11
\long\def\@makefntext#1{
\protect\noindent \hbox to 3.2pt {\hskip-.9pt  
$^{{\eightrm\@thefnmark}}$\hfil}#1\hfill}		

\def\thefootnote{\fnsymbol{footnote}}
\def\@makefnmark{\hbox to 0pt{$^{\@thefnmark}$\hss}}	
	
\def\ps@myheadings{\let\@mkboth\@gobbletwo
\def\@oddhead{\hbox{}
\rightmark\hfil\eightrm\thepage}   
\def\@oddfoot{}\def\@evenhead{\eightrm\thepage\hfil
\leftmark\hbox{}}\def\@evenfoot{}
\def\sectionmark##1{}\def\subsectionmark##1{}}

\oddsidemargin=\evensidemargin
\renewcommand{\thefootnote}{\fnsymbol{footnote}}

\newcounter{sectionc}\newcounter{subsectionc}\newcounter{subsubsectionc}
\renewcommand{\section}[1] {\vspace{12pt}\addtocounter{sectionc}{1} 
\setcounter{subsectionc}{0}\setcounter{subsubsectionc}{0}\noindent 
	{\tenbf\thesectionc. #1}\par\vspace{5pt}}
\renewcommand{\subsection}[1] {\vspace{12pt}\addtocounter{subsectionc}{1} 
	\setcounter{subsubsectionc}{0}\noindent 
	{\bf\thesectionc.\thesubsectionc. {\kern1pt \bfit #1}}\par\vspace{5pt}}
\renewcommand{\subsubsection}[1] {\vspace{12pt}\addtocounter{subsubsectionc}{1}
	\noindent{\tenrm\thesectionc.\thesubsectionc.\thesubsubsectionc.
	{\kern1pt \tenit #1}}\par\vspace{5pt}}
\newcommand{\nonumsection}[1] {\vspace{12pt}\noindent{\tenbf #1}
	\par\vspace{5pt}}

\newcounter{appendixc}
\newcounter{subappendixc}[appendixc]
\newcounter{subsubappendixc}[subappendixc]
\renewcommand{\thesubappendixc}{\Alph{appendixc}.\arabic{subappendixc}}
\renewcommand{\thesubsubappendixc}
	{\Alph{appendixc}.\arabic{subappendixc}.\arabic{subsubappendixc}}

\renewcommand{\appendix}[1] {\vspace{12pt}
        \refstepcounter{appendixc}
        \setcounter{figure}{0}
        \setcounter{table}{0}
        \setcounter{lemma}{0}
        \setcounter{theorem}{0}
        \setcounter{corollary}{0}
        \setcounter{definition}{0}
        \setcounter{equation}{0}
        \renewcommand{\thefigure}{\Alph{appendixc}.\arabic{figure}}
        \renewcommand{\thetable}{\Alph{appendixc}.\arabic{table}}
        \renewcommand{\theappendixc}{\Alph{appendixc}}
        \renewcommand{\thelemma}{\Alph{appendixc}.\arabic{lemma}}
        \renewcommand{\thetheorem}{\Alph{appendixc}.\arabic{theorem}}
        \renewcommand{\thedefinition}{\Alph{appendixc}.\arabic{definition}}
        \renewcommand{\thecorollary}{\Alph{appendixc}.\arabic{corollary}}
        \renewcommand{\theequation}{\Alph{appendixc}.\arabic{equation}}
        \noindent{\tenbf Appendix \theappendixc #1}\par\vspace{5pt}}
\newcommand{\subappendix}[1] {\vspace{12pt}
        \refstepcounter{subappendixc}
        \noindent{\bf Appendix \thesubappendixc. {\kern1pt \bfit #1}}
	\par\vspace{5pt}}
\newcommand{\subsubappendix}[1] {\vspace{12pt}
        \refstepcounter{subsubappendixc}
        \noindent{\rm Appendix \thesubsubappendixc. {\kern1pt \tenit #1}}
	\par\vspace{5pt}}

\newcommand{\textlineskip}{\baselineskip=13pt}
\newcommand{\smalllineskip}{\baselineskip=10pt}

\def\eightcirc{
\begin{picture}(0,0)
\put(4.4,1.8){\circle{6.5}}
\end{picture}}
\def\eightcopyright{\eightcirc\kern2.7pt\hbox{\eightrm c}} 

\newcommand{\copyrightheading}[1]
	{\vspace*{-2.5cm}\smalllineskip{\flushleft
	{\footnotesize International Journal of Modern Physics A, #1}\\
	{\footnotesize $\eightcopyright$\, World Scientific Publishing
	 Company}\\
	 }}

\def\abstracts#1#2#3{{
	\centering{\begin{minipage}{4.5in}\baselineskip=10pt\footnotesize
	\parindent=0pt #1\par 
	\parindent=15pt #2\par
	\parindent=15pt #3
	\end{minipage}}\par}} 

\newcommand{\bibit}{\nineit}

\renewenvironment{thebibliography}[1]
	{\frenchspacing
	 \ninerm\baselineskip=11pt
	 \begin{list}{\arabic{enumi}.}
	{\usecounter{enumi}\setlength{\parsep}{0pt}
	 \setlength{\leftmargin 12.7pt}{\rightmargin 0pt} 
	 \setlength{\itemsep}{0pt} \settowidth
	{\labelwidth}{#1.}\sloppy}}{\end{list}}

\newcounter{itemlistc}
\newcounter{romanlistc}
\newcounter{alphlistc}
\newcounter{arabiclistc}

\newcommand{\fcaption}[1]{
        \refstepcounter{figure}
        \setbox\@tempboxa = \hbox{\footnotesize Fig.~\thefigure. #1}
        \ifdim \wd\@tempboxa > 5in
           {\begin{center}
        \parbox{5in}{\footnotesize\smalllineskip Fig.~\thefigure. #1}
            \end{center}}
        \else
             {\begin{center}
             {\footnotesize Fig.~\thefigure. #1}
              \end{center}}
        \fi}

\newcommand{\tcaption}[1]{
        \refstepcounter{table}
        \setbox\@tempboxa = \hbox{\footnotesize Table~\thetable. #1}
        \ifdim \wd\@tempboxa > 5in
           {\begin{center}
        \parbox{5in}{\footnotesize\smalllineskip Table~\thetable. #1}
            \end{center}}
        \else
             {\begin{center}
             {\footnotesize Table~\thetable. #1}
              \end{center}}
        \fi}

\def\@citex[#1]#2{\if@filesw\immediate\write\@auxout
	{\string\citation{#2}}\fi
\def\@citea{}\@cite{\@for\@citeb:=#2\do
	{\@citea\def\@citea{,}\@ifundefined
	{b@\@citeb}{{\bf ?}\@warning
	{Citation `\@citeb' on page \thepage \space undefined}}
	{\csname b@\@citeb\endcsname}}}{#1}}

\newif\if@cghi
\def\cite{\@cghitrue\@ifnextchar [{\@tempswatrue
	\@citex}{\@tempswafalse\@citex[]}}
\def\citelow{\@cghifalse\@ifnextchar [{\@tempswatrue
	\@citex}{\@tempswafalse\@citex[]}}
\def\@cite#1#2{{$\null^{#1}$\if@tempswa\typeout
	{IJCGA warning: optional citation argument 
	ignored: `#2'} \fi}}

\def\pmb#1{\setbox0=\hbox{#1}
	\kern-.025em\copy0\kern-\wd0
	\kern.05em\copy0\kern-\wd0
	\kern-.025em\raise.0433em\box0}

\def\fnt#1#2{\footnotetext{\kern-.3em
	{$^{\mbox{\scriptsize #1}}$}{#2}}}

\def\fpage#1{\begingroup
\voffset=.3in
\thispagestyle{empty}\begin{table}[b]\centerline{\footnotesize #1}
	\end{table}\endgroup}

\def\runninghead#1#2{\pagestyle{myheadings}
\markboth{{\protect\footnotesize\it{\quad #1}}\hfill}
{\hfill{\protect\footnotesize\it{#2\quad}}}}
\font\tenrm=cmr10
\font\tenit=cmti10 
\font\tenbf=cmbx10
\font\bfit=cmbxti10 at 10pt
\font\ninerm=cmr9
\font\nineit=cmti9

\font\eightrm=cmr8

\def\qed{\hbox{${\vcenter{\vbox{			
   \hrule height 0.4pt\hbox{\vrule width 0.4pt height 6pt
   \kern5pt\vrule width 0.4pt}\hrule height 0.4pt}}}$}}

\renewcommand{\thefootnote}{\fnsymbol{footnote}}	

\newcommand{\epjc}[3]{{\bibit Eur. Phys. J.} {\bf C#1} (#2) #3.}

\newcommand{\pn}[3]{The OPAL Collaboration, {\bibit #3}, OPAL Physics Note 
#1, #2.}

\begin{document}

\runninghead{Search for R-parity Violating Decays of Supersymmetric Particles at
LEP}{Search for R-parity Violating Decays of Supersymmetric Particles at LEP}

\normalsize\textlineskip
\thispagestyle{empty}
\setcounter{page}{1}

\copyrightheading{}         

\vspace*{-7mm}

\begin{flushright}
OPAL Conference Report CR452 \\ 20 November 2000
\end{flushright}		

\vspace*{-7mm}

\vspace*{0.88truein}

\fpage{1}
\centerline{\bf SEARCH FOR R-PARITY VIOLATING DECAYS}
\centerline{\bf OF SUPERSYMMETRIC PARTICLES AT LEP}
\vspace*{0.37truein}
\centerline{\footnotesize GABRIELLA P\'ASZTOR\footnote{
Permanent address: KFKI Research Institute for Particle and Nuclear Physics of
the Hungarian Academy of Sciences (KFKI RMKI), Budapest, P.O.Box 49, H-1525, 
Hungary.
Supported partially by the Hungarian Foundation of Scientific Research under 
the contract number OTKA F-023259.}}
\vspace*{0.015truein}
\centerline{\footnotesize\it CERN EP}
\baselineskip=10pt
\centerline{\footnotesize\it Geneva 23, CH-1211, Switzerland}

\vspace*{0.21truein}
\abstracts{Searches for pair-produced charginos, neutralinos and scalar fermions
decaying via R-parity violating $\lambda, \lambda^\prime$ and
$\lambda^{\prime\prime}$ couplings with the OPAL detector at LEP are presented
at $\sqrt{s}=189$ GeV. Partial updates using data up to the highest energies 
of LEP, $\sqrt{s}=209$ GeV, are also given.}{}{}

\textlineskip			
\vspace*{12pt}			

\noindent
If supersymmetry (SUSY) is the answer to the hierarchy problem of the Standard 
Model (SM), there should be a superpartner to each SM particle. We present here
a search for pair-produced supersymmetric particles in e$^+$e$^-$ collisions
with the OPAL detector, assuming that R-parity can be violated. 


R-parity violating (RPV) interactions are 
parametrized with a gauge-invariant superpotential that includes the 
following Yukawa coupling terms: 
\begin{eqnarray}
{\cal W}_{RPV}  = 
    \lambda_{ijk}      L_i L_j {\overline E}_k
 +  \lambda^{'}_{ijk}  L_i Q_j {\overline D}_k
 +  \lambda^{''}_{ijk} {\overline U}_i {\overline D}_j {\overline D}_k, 
\label{lagrangian}
\end{eqnarray}
where $i,j,k$ are the generation indices of the superfields 
$L, Q,E,D$ and $U$. $L$ and $Q$ are lepton and quark left-handed doublets,  
while
$\overline E$, $\overline D$ and $\overline U$ are right-handed 
singlet charge-conjugate superfields for the charged 
leptons, down- and up-type quarks, respectively. 
There are nine $\lambda$, 27 $\lambda^\prime$
and nine $\lambda^{\prime\prime}$
couplings, giving a total of 45
$\lambda$-like R-parity violating couplings.  

R-parity is a discreet, multiplicative quantum number, which is +1 for SM
particles and $-1$ for their superpartners. There is no {\it a priori} law, that
requires the conservation of R-parity. Moreover there is no experimental result,
which excludes the presence of ${\cal W}_{RPV}$ under the assumption that only
one $\lambda$-like coupling is significantly different from zero.
 
The model used in this work is the Constrained Minimal Supersymmetric
Standard Model (CMSSM).  
Not counting the 45 RPV Yukawa couplings,
this model has only five free parameters: 
a common mass for the gauginos ($m_{1/2}$) and  
the sfermions ($m_0$) at the GUT scale,
the mixing parameter of the two Higgs field doublets ($\mu$),
the ratio of the vacuum expectation values of the two Higgs doublets
($\tan\beta$) and the common trilinear coupling ($A$). 

Both the theoretical results and the details of the experimental 
analyses relevant for this paper are summarized in our previous 
publications.\cite{rpv-sfermion,rpv-gaugino} We assume, that there is only
one $\lambda$-like coupling different from zero, and that SUSY particles are heavy
and decay promptly in the detector. This corresponds to a sensitivity for
couplings larger than ${\cal O}(10^{-5})$.

The bulk of the results,\cite{rpv-lambda,rpv-lambdap} presented in Section 1 to
3, are based on the combination of the data sample collected at $\sqrt{s}=189$
GeV corresponding to an integrated luminosity of approximately 180 pb$^{-1}$
with previous OPAL results.\cite{rpv-sfermion,rpv-gaugino} Partial updates 
using data collected up to $\sqrt{s}=209$ are presented Section 4.

\textheight=7.8truein
\setcounter{footnote}{0}
\renewcommand{\thefootnote}{\alph{footnote}}

\section{Chargino and Neutralino Searches}
\noindent
The direct decays of neutralinos result in three SM fermions:
$\nu\ell^+\ell^-$ via $\lambda$, $\nu$qq or $\ell$qq via
$\lambda^\prime$ and qqq via $\lambda^{\prime\prime}$ couplings.
Thus, the final state topologies of their pair-production
vary from four leptons with 
missing energy to six jets.

Charginos can decay directly to SM fermions, or indirectly through a neutralino
and a (virtual) W boson. In direct decays, three SM fermions are produced:
$\nu\nu\ell$ or $\ell\ell\ell$ via $\lambda$, $\nu$qq or $\ell$qq via
$\lambda^\prime$ and qqq via $\lambda^{\prime\prime}$ couplings.
When decaying indirectly, five SM fermions are created, their type
depending on the decay modes of the neutralino and the W boson.
Therefore, pair-production of charginos may be observed in widely different
final states, ranging from two leptons with missing energy to ten jets.

When interpreting the results of all these topological searches, first
cross-section limits are derived for the individual channels with the
minimal model assumptions given above.
We combine the individual results to get cross-section limits for the direct
and the indirect modes. It is assumed that the indirect decays proceed through
the production of a (virtual) W boson and the lightest neutralino, which is
assumed to be the lightest SUSY particle, and that the mass difference
between the chargino and the neutralino is greater than 5 GeV. Different
channels are combined according to the measured W branching ratios.

Decay mode independent results are calculated by varying the relative
branching ratio of the direct and indirect decays between 0 and 1, and taking
 the worst case.

For any $\lambda$ coupling, the upper limit on the neutralino pair-production
cross-section is smaller than 0.16 pb in the full mass range from
45 GeV to the kinematic limit. The cross-section limits for charginos are
summarized in Table~1.

\begin{table}[h!]
\tcaption{Upper limits on the chargino pair-production cross-section
for direct decays, indirect decays and independently of the decay mode
for any $\lambda$, $\lambda^\prime$ and $\lambda^{\prime\prime}$ couplings.
The first / second number corresponds to a chargino mass of 45 / 90 GeV.}
\centerline{\footnotesize\smalllineskip
\begin{tabular}{cccc}\\
\hline
coupling & direct & indirect & mode indep. \\
\hline
$\lambda$ & 1.1 / 0.55 pb & 0.42 / 0.16 pb & 2.0 / 0.9 pb \\
$\lambda^\prime$ & 2.0 / 0.35 pb & 4.5 / 0.7 pb & \\
$\lambda^{\prime\prime}$ & 1.4 / 0.50 pb & 1.1 / 0.50 pb & 1.4 /0.55 pb \\
\hline
\end{tabular}}
\end{table}

Finally, the results are interpreted in CMSSM. The excluded topological
cross-sections, and the measured Z boson width are used to
constrain the CMSSM parameter space. The results for any $\lambda$ and
$\lambda^{\prime\prime}$ couplings are shown in Figure 1.

\begin{figure}[htbp]
\begin{center}
{\bf OPAL Preliminary}\hspace*{3cm}{\bf OPAL Preliminary}
\end{center}

\includegraphics*[scale=0.35]{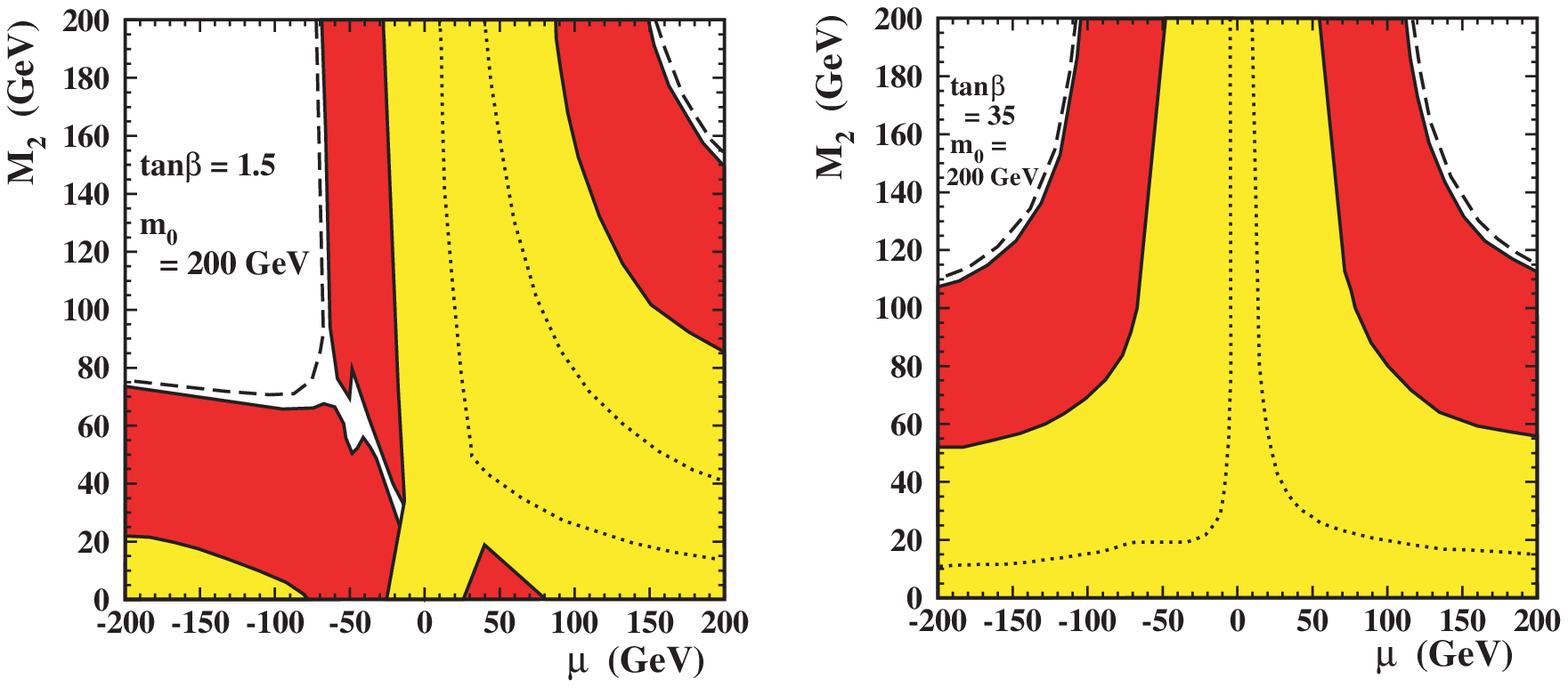} \hfill 
\includegraphics*[scale=0.16]{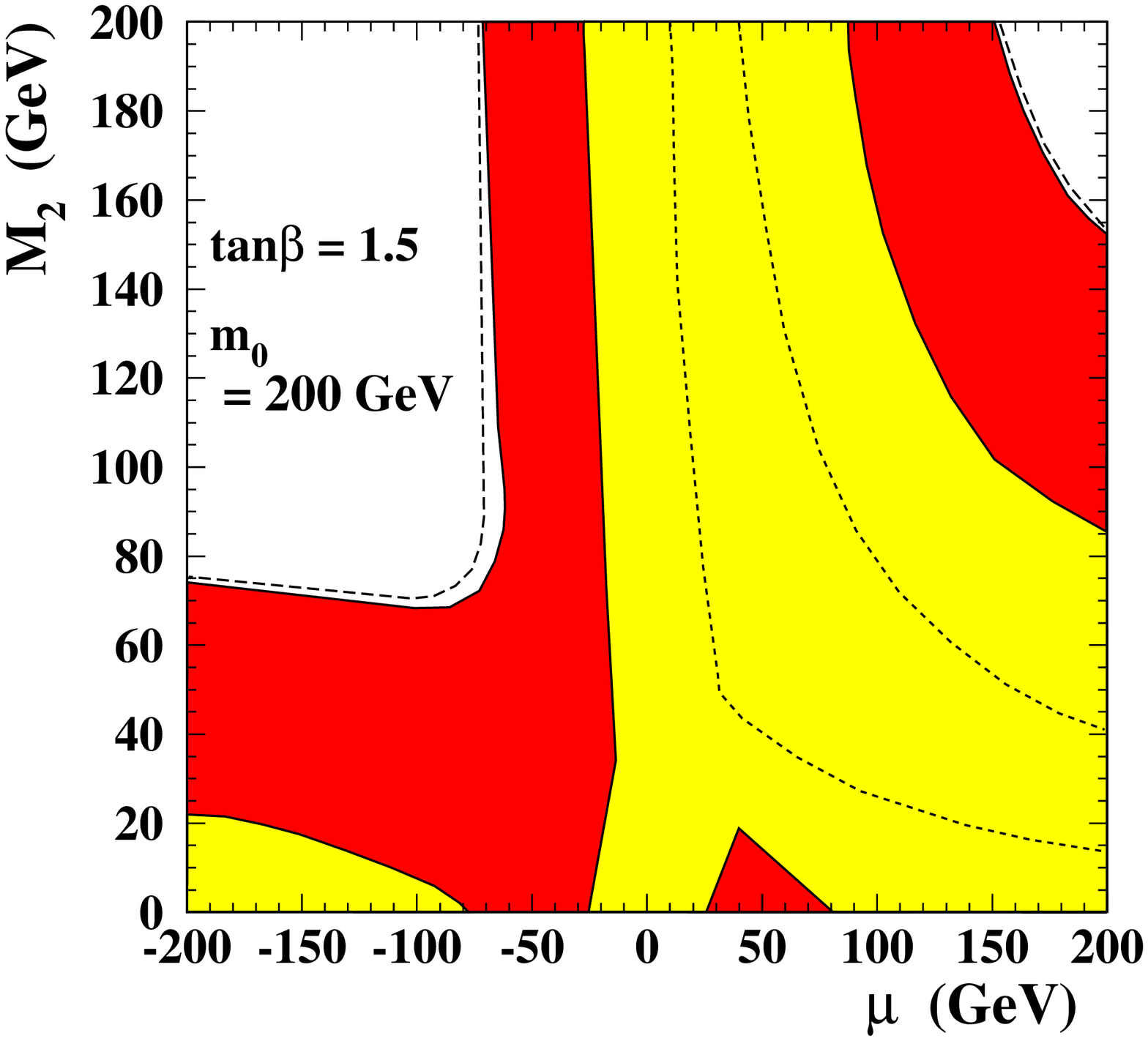} 
\includegraphics*[scale=0.16]{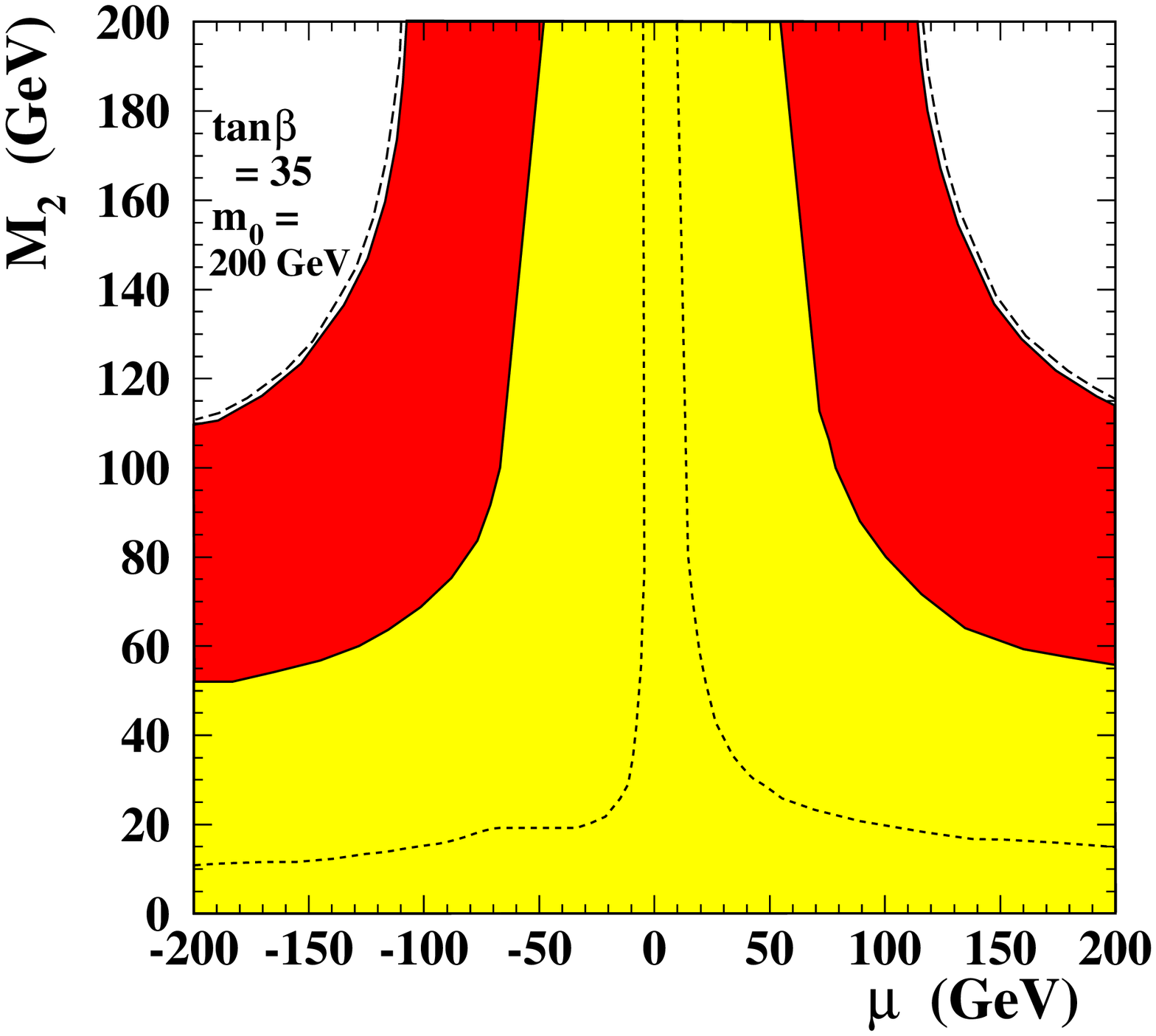} 
\vspace*{-3.1cm}

{\small 
\hspace*{2.45cm}(a)\hspace*{2.8cm}(b)\hspace*{2.8cm}(c)\hspace*{2.8cm}(d)
\vspace*{1cm}
}

\hspace*{2.35cm}$_{\lambda}$\hspace*{2.88cm}$_{\lambda}$\hspace*{2.88cm}
$_{\lambda^{\prime\prime}}$\hspace*{2.88cm}$_{\lambda^{\prime\prime}}$
\vspace*{0.95cm}

\fcaption{Excluded regions in CMSSM in the $M_2 - \mu$ plane$^a$ for 
$m_0=200$ GeV and (a,c) $\tan\beta=1.5$, (b,d) $\tan\beta=35$ 
for any (a,b) $\lambda$ and
(c,d) $\lambda^{\prime\prime}$ couplings. The kinematic limit is shown by 
dashed lines.}
\vspace*{-15pt}
\end{figure}

\footnotetext{$^a$ $M_2$, the SU(2) soft SUSY breaking gaugino mass, 
is related to $m_{1/2}$.}

\section{Squark Searches}
\noindent
We have studied the direct decays of scalar top quarks to SM fermions
via $\lambda^\prime$ and $\lambda^{\prime\prime}$ couplings. The cross-section
upper limits are shown for both the q$\ell$q$\ell$ and the qqqq final states 
on Figure 2. In CMSSM the
production cross-section depends on the stop mixing angle,
$\theta_{\tilde{t}}$, defined as $\tilde{t}_1 = \cos\theta_{\tilde{t}} 
\tilde{t}_L + 
\sin\theta_{\tilde{t}} \tilde{t}_R$. The maximal cross-section is predicted for 
$\theta_{\tilde{t}}=0$ rad, while the minimal for $\theta_{\tilde{t}}=0.98$ rad.
The limit on the stop mass varies between 79 and
90 GeV depending on which $\lambda$-like coupling is different from zero
and the value of the mixing angle. 

\begin{figure}[htbp]
\vspace*{-10pt}
\centering
\includegraphics*[height=3.5cm,width=6.2cm]{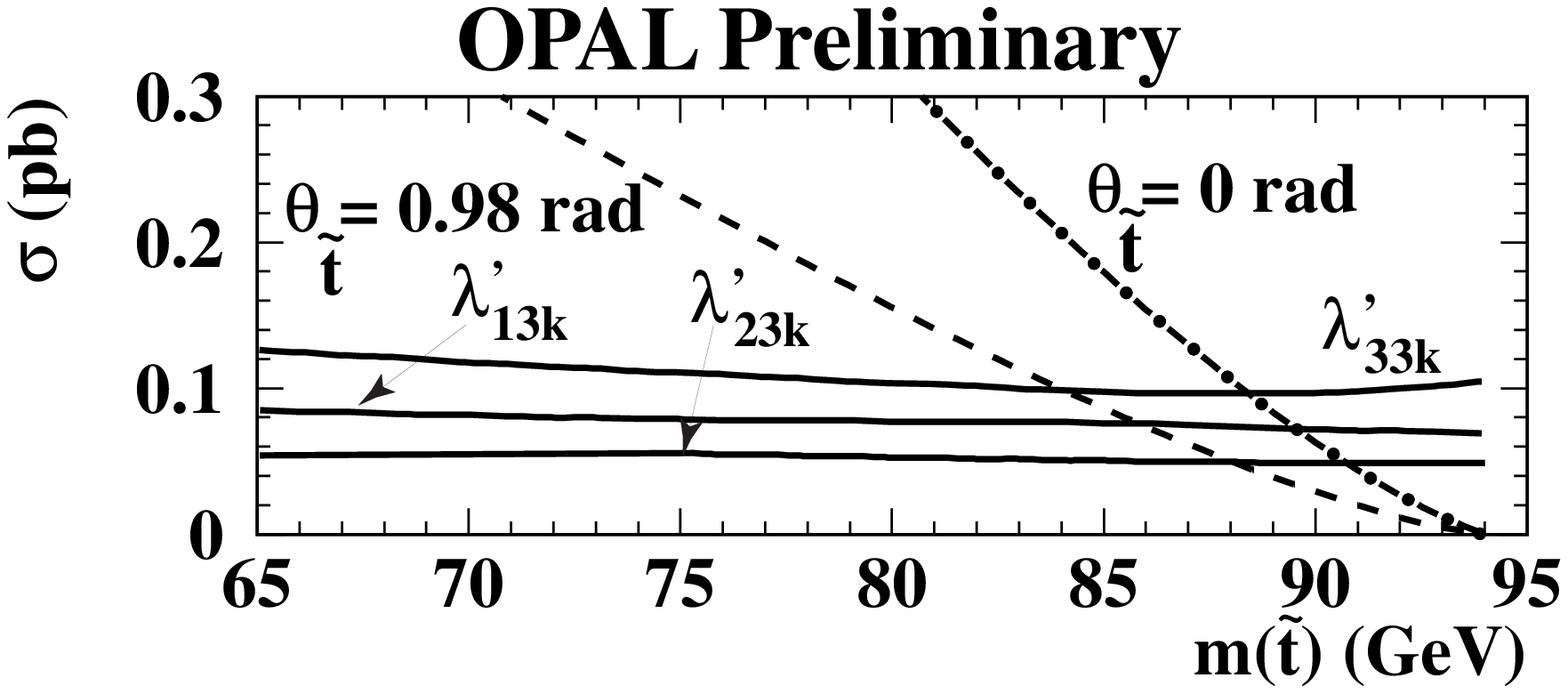} 
\includegraphics*[height=3.5cm,width=6.2cm]{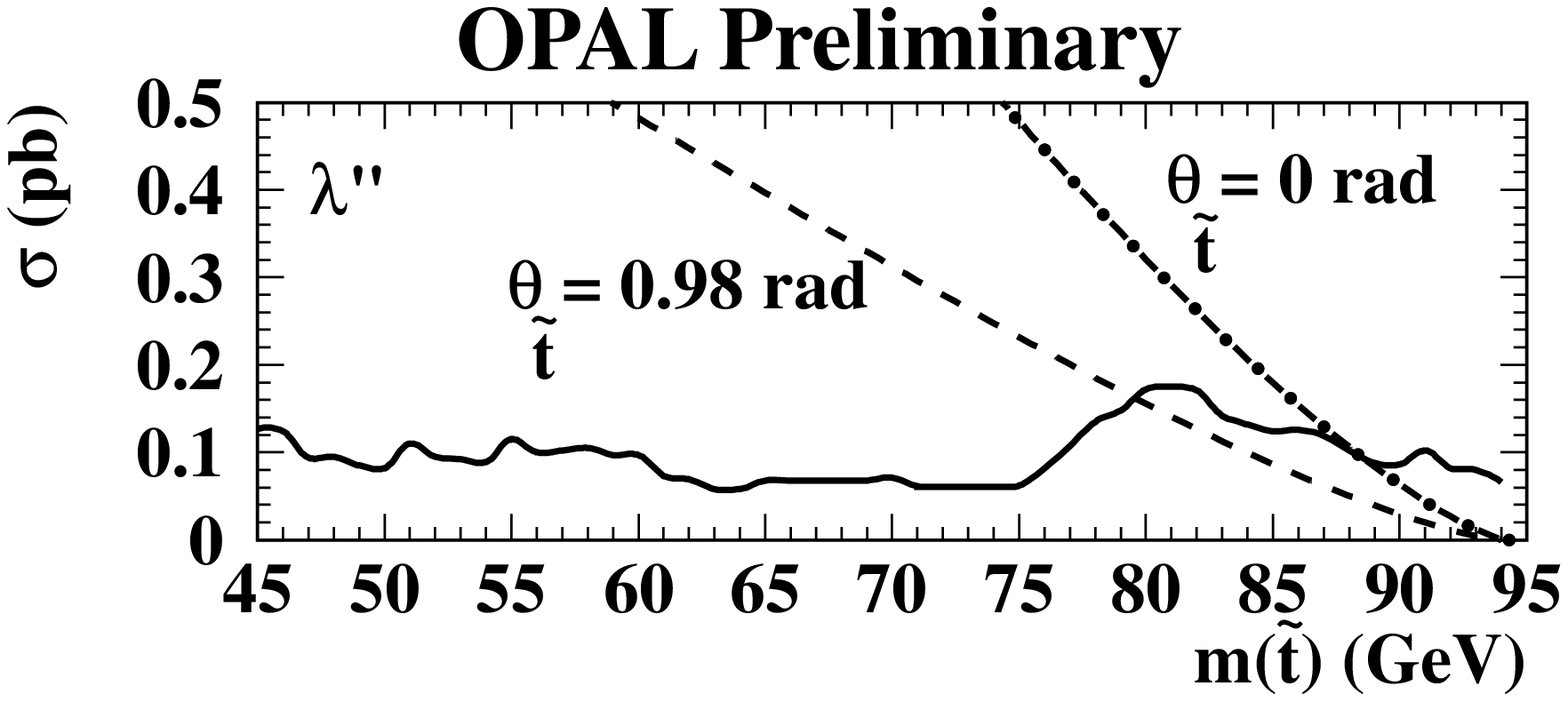} 
\vspace*{-2.5cm}

\hspace*{4.8cm}(a)\hspace*{5.8cm}(b)
\vspace*{2cm}

\fcaption{Upper limits on the pair-production cross-section of (a) scalar top
quarks decaying via $\lambda^\prime$ couplings and (b) a scalar quark of any
flavor decaying via $\lambda^{\prime\prime}$ couplings. The predicted
minimal and maximal cross-sections of stop pair-production
in CMSSM are also shown.}
\vspace*{-15pt}
\end{figure}

\section{Slepton Searches}
\noindent
Scalar leptons, charged or neutral, can decay directly to two SM fermions or
indirectly, through the production of a SM lepton and a neutralino, to four SM
fermions. Thus, when produced in pairs, the possible final states range from two
leptons and missing energy to four leptons and four hadronic jets.

Similarly to the gaugino searches, first cross-section limits are computed
with minimal model assumptions. Then a lower limit on the slepton mass is
derived in CMSSM. In the case of direct decays, it is assumed that the
branching ratio $BR(\tilde{f}_i \to f_j f_k) = 1$. For indirect decays the
branching ratio $BR(\tilde{f} \to f \tilde{\chi}^0)$ is taken from CMSSM. 
Since the neutralino decay modes
via $\lambda^\prime$ couplings
depend on CMSSM parameters, we vary the branching ratio of the decay involving a
charged lepton and the branching ratio of the decay involving a neutrino
simultaneously between 0 and 1, and take the worst case result.


As previously, only the weakest limits are shown. 
All CMSSM exclusion
plots are given for $\tan\beta = 1.5$ and $\mu=-200$ GeV. This choice of
parameters is conservative, since the theoretical cross-section usually
increases
for larger $\tan\beta$ and $|\mu |$.

The upper limit on the sneutrino pair-production cross-section is better than
0.22 pb in the mass range of interest for any $\lambda$ coupling. The
cross-section limits are generally weaker for $\lambda^\prime$ couplings. For
example, for indirect decays of sneutrinos, 
the limit is 1.2$-$1.3 pb for a mass of 45 GeV and 0.2 pb for a mass of
90 GeV. The CMSSM mass limits are shown in Figure 3(a) and (b) for any
$\lambda$ and $\lambda^\prime$ couplings.

\begin{figure}[htbp]
\centering
{\bf OPAL preliminary \hspace*{2.5cm} OPAL preliminary}
\vspace*{-5mm}

\noindent
\includegraphics*[scale=0.35,clip]{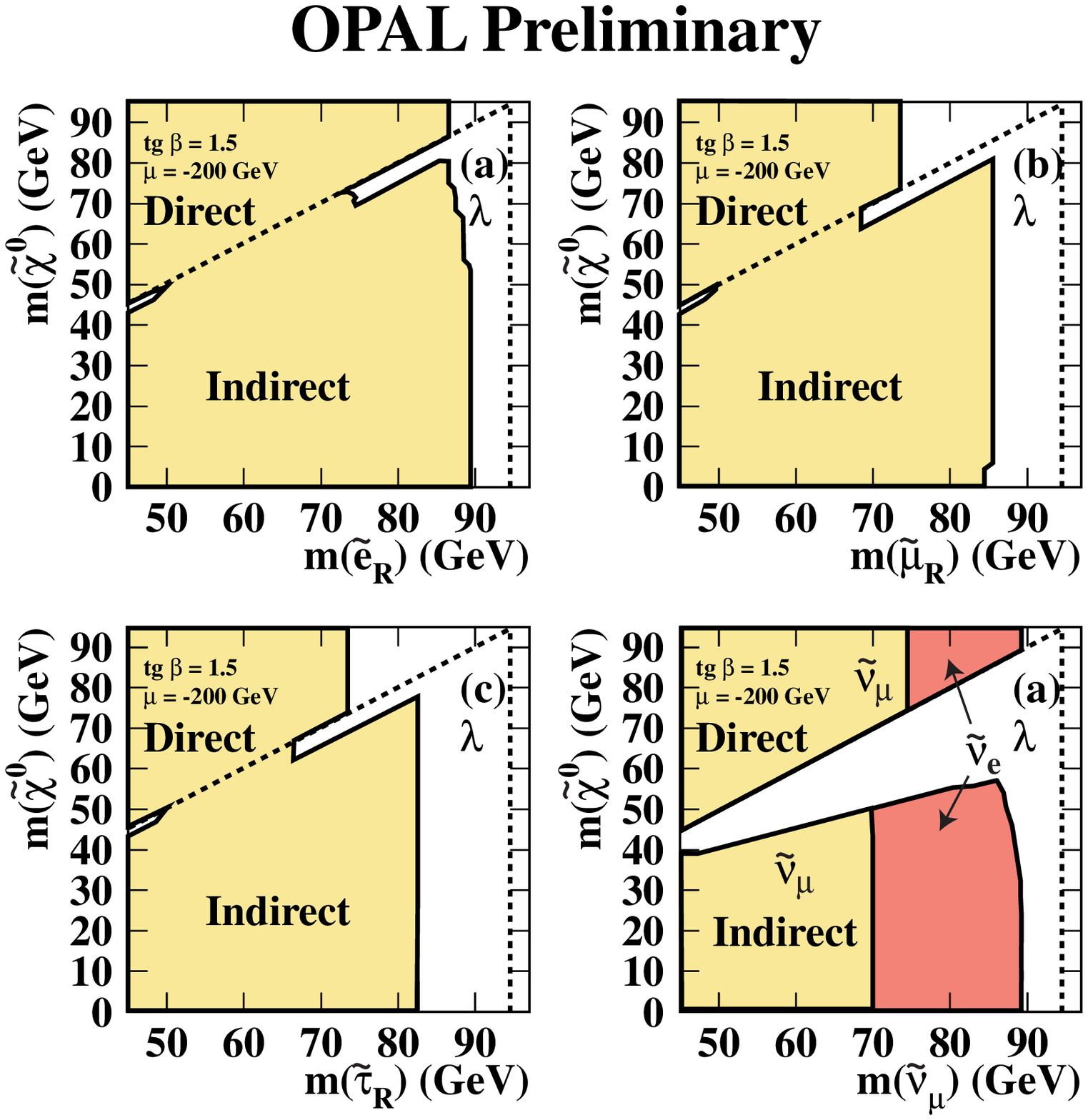} 
\includegraphics*[scale=0.155]{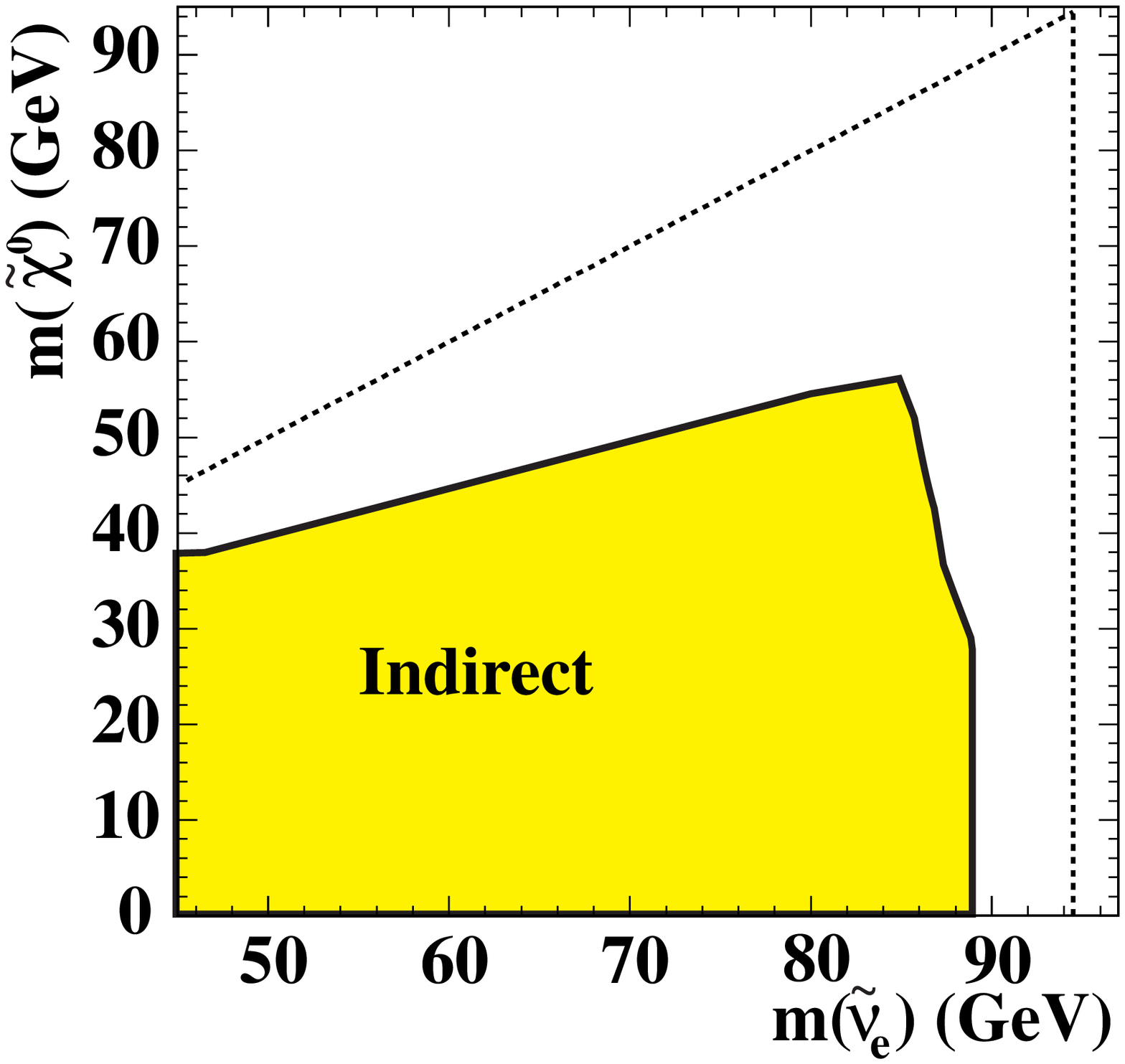} 
\includegraphics*[scale=0.35]{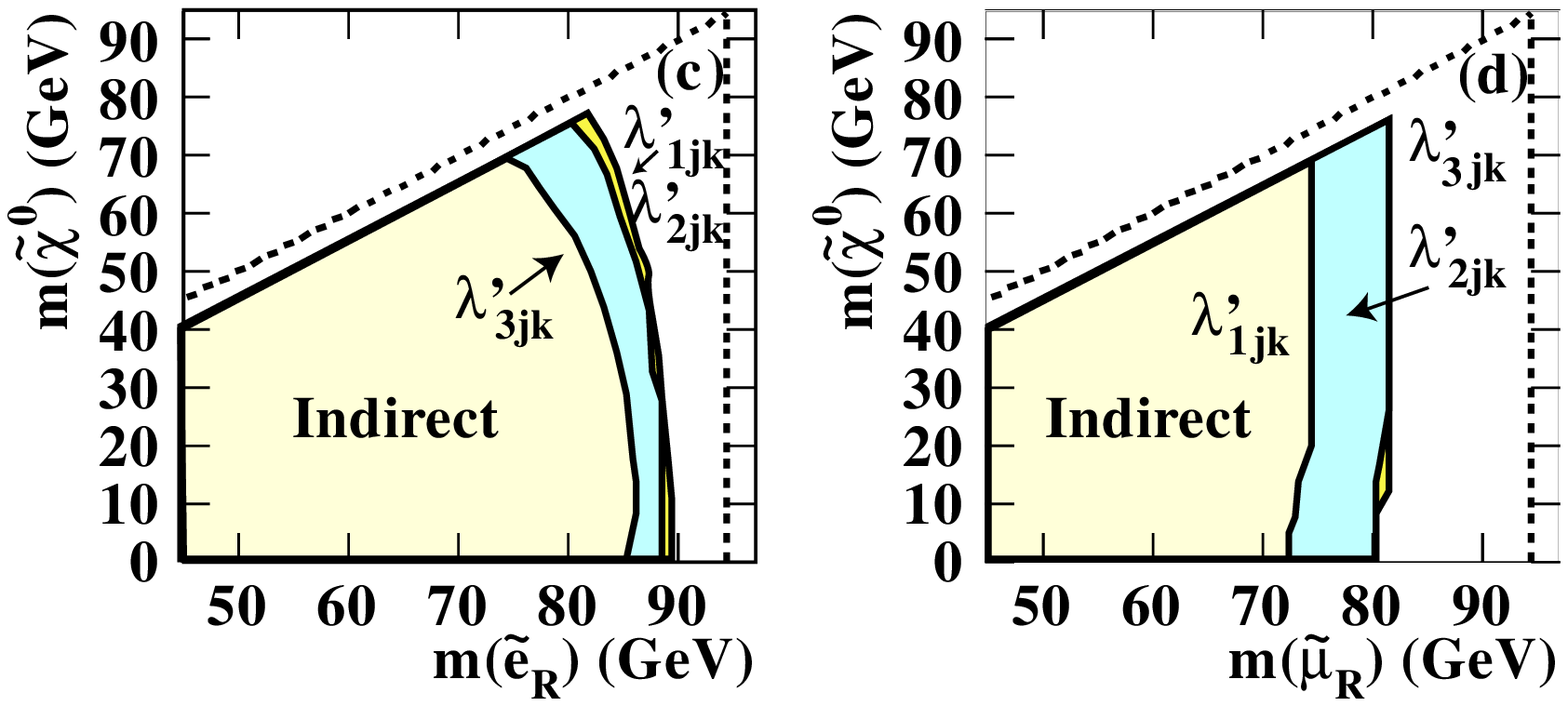} 
\vspace*{-2.7cm}

\hspace*{-1cm}{\footnotesize (b)}
\vspace*{2.2cm}

\fcaption{Excluded mass regions in CMSSM (a) from direct and indirect decays
of sneutrinos for any $\lambda$ coupling, and
from indirect decays of (b) electron sneutrinos, (b) right-handed
selectrons and (c) right-handed smuons for any
$\lambda^\prime$ coupling. No CMSSM exclusion is possible for
muon and tau sneutrinos decaying via $\lambda^\prime$ couplings
due to the small expected theoretical cross-section.
The kinematic limit is shown by dashed lines.}
\end{figure}

The charged slepton limits for $\lambda$ couplings have been updated using data
collected at  $\sqrt{s}=192-209$ and are summarized in the next section. 
For $\lambda^\prime$ couplings, the selectron and smuon 
pair-production cross-section upper limits for indirect
decays are better than 0.52 pb.
The CMSSM mass limits are given in Figure 3 (c,d).

\boldmath 
\section{Updates at $\sqrt{s}=192-209$ GeV}   
\unboldmath 
\noindent 
Preliminary updates\cite{search2000nov} using data collected in 1999 and 2000,
corresponding to an integrated luminosity of 400 pb$^{-1}$, are performed in
the multi-lepton and in the  jets+leptons final states. There is a nice
agreement between data and the SM expectation, with the exception of events with 
multiple jets and identified tau leptons at the highest energy, 
$\sqrt{s}>206.5$
GeV. The biggest excess, corresponding to a Poisson probability of
$2\times10^{-3}$, is found in the final state with two tau leptons and at least
four jets, in the search
for indirect decays of gauginos via $\lambda^\prime$ coupling, with 8 events
observed and 2.2 expected from SM processes.

Including around 285 pb$^{-1}$ of $\sqrt{s}=192-209$ GeV data collected by July
2000, new upper limits, shown in Figure 4(a), are set on the production
cross-section of charged sleptons decaying via $\lambda$
couplings.\cite{search2000july} Limits for direct decays are
weaker, the cross-section limit, based on 200 pb$^{-1}$ data collected at 
$\sqrt{s}=192-202$ GeV and combined with previous results, is
shown in Figure 5(a).\cite{search1999nov} 

\begin{figure}[htbp]
\centering
\includegraphics*[width=5cm,height=4cm]{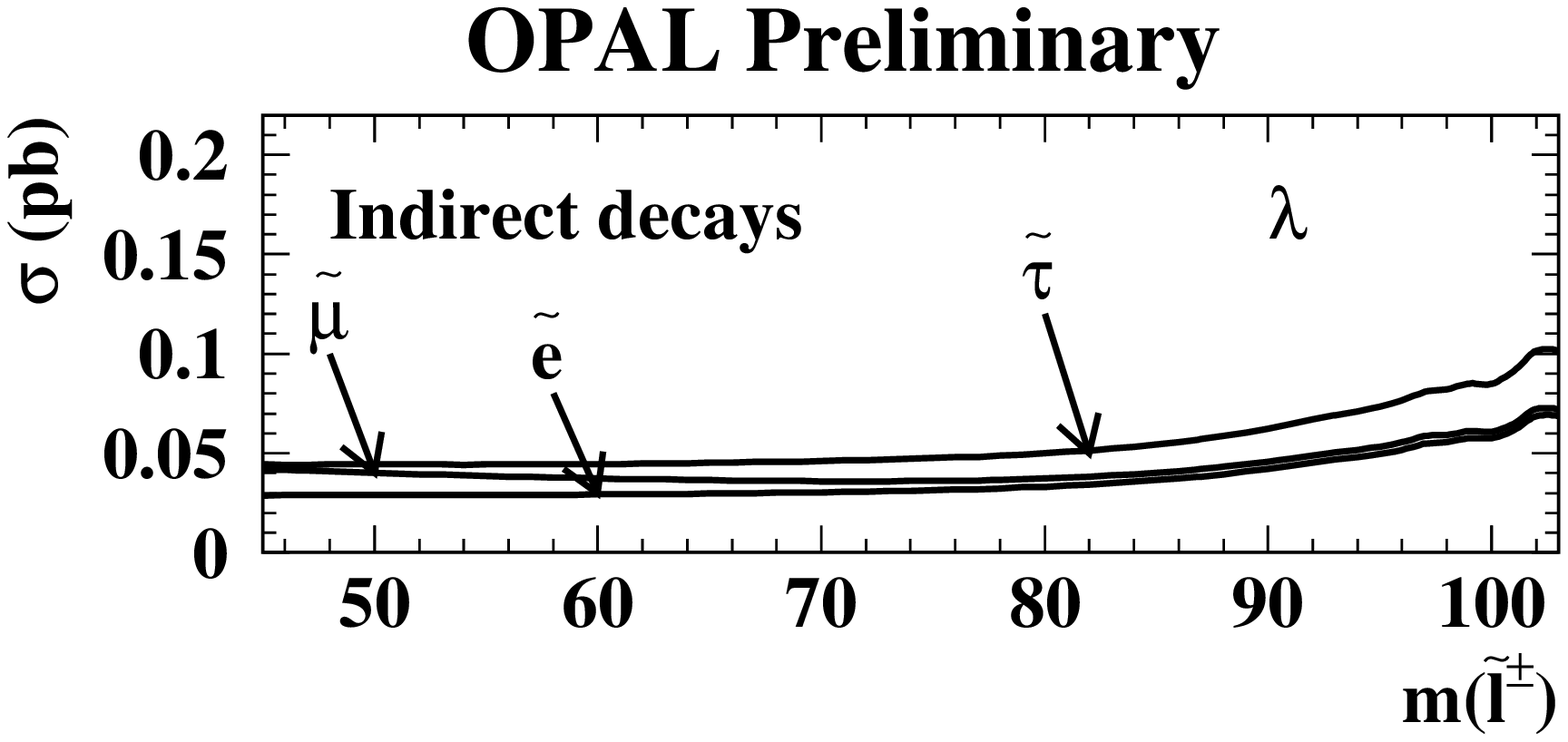} 
\includegraphics*[width=7.5cm,height=4cm]{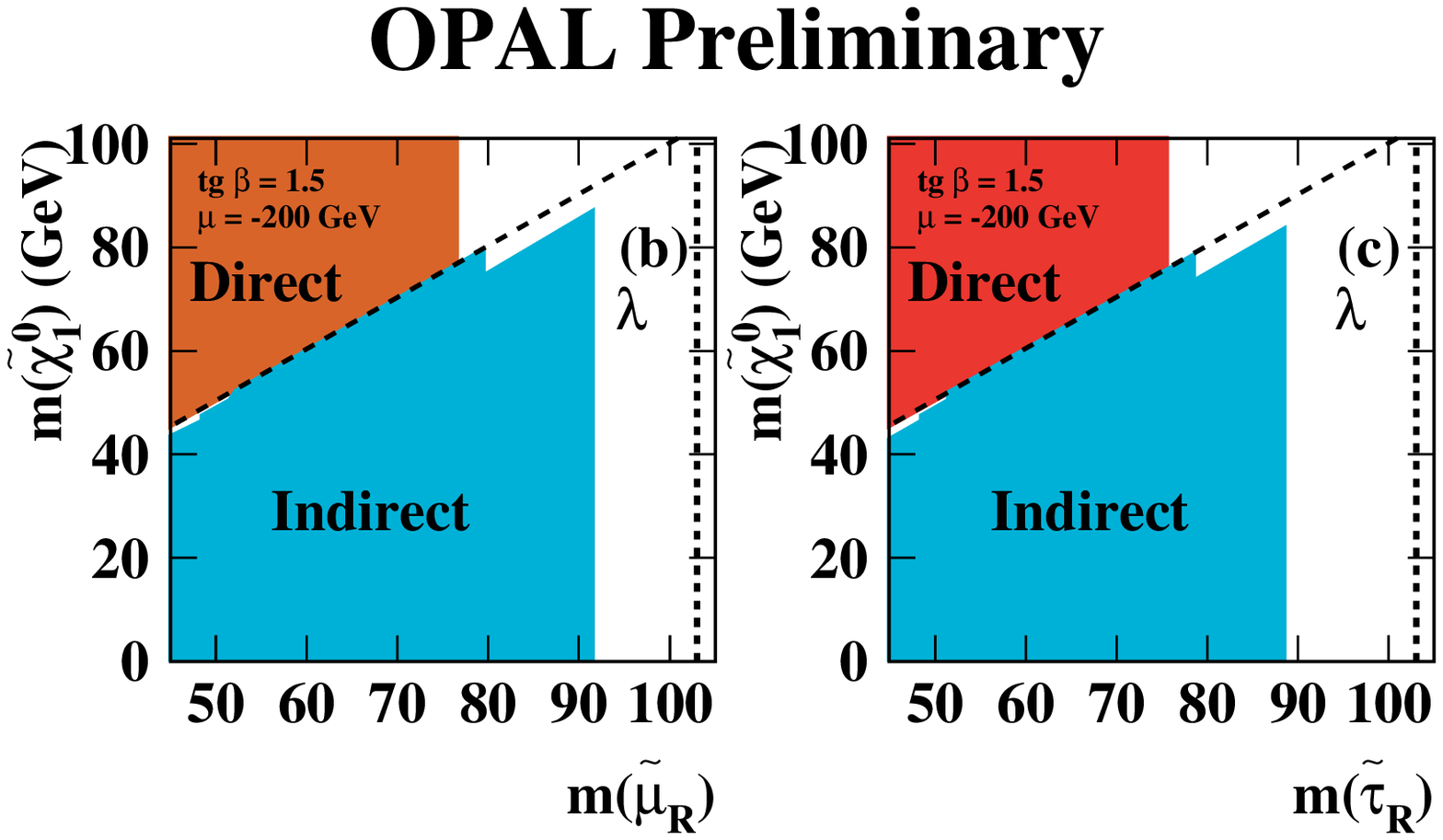} 
\vspace*{-3.65cm}

\hspace*{-3.5cm} (a)
\vspace*{2.8cm}
\fcaption{(a) Upper limit on the
pair-production cross-sections for indirect decays of
$\tilde{\ell}_{\mathrm{R}}$. CMSSM exclusion regions for (b) smuon and
(c) stau pair-production. The kinematic limit is shown by dashed lines.}
\end{figure}

\begin{figure}[htbp]
\vspace*{-15pt}
\begin{center}
\includegraphics*[height=3.7cm,width=6.2cm]{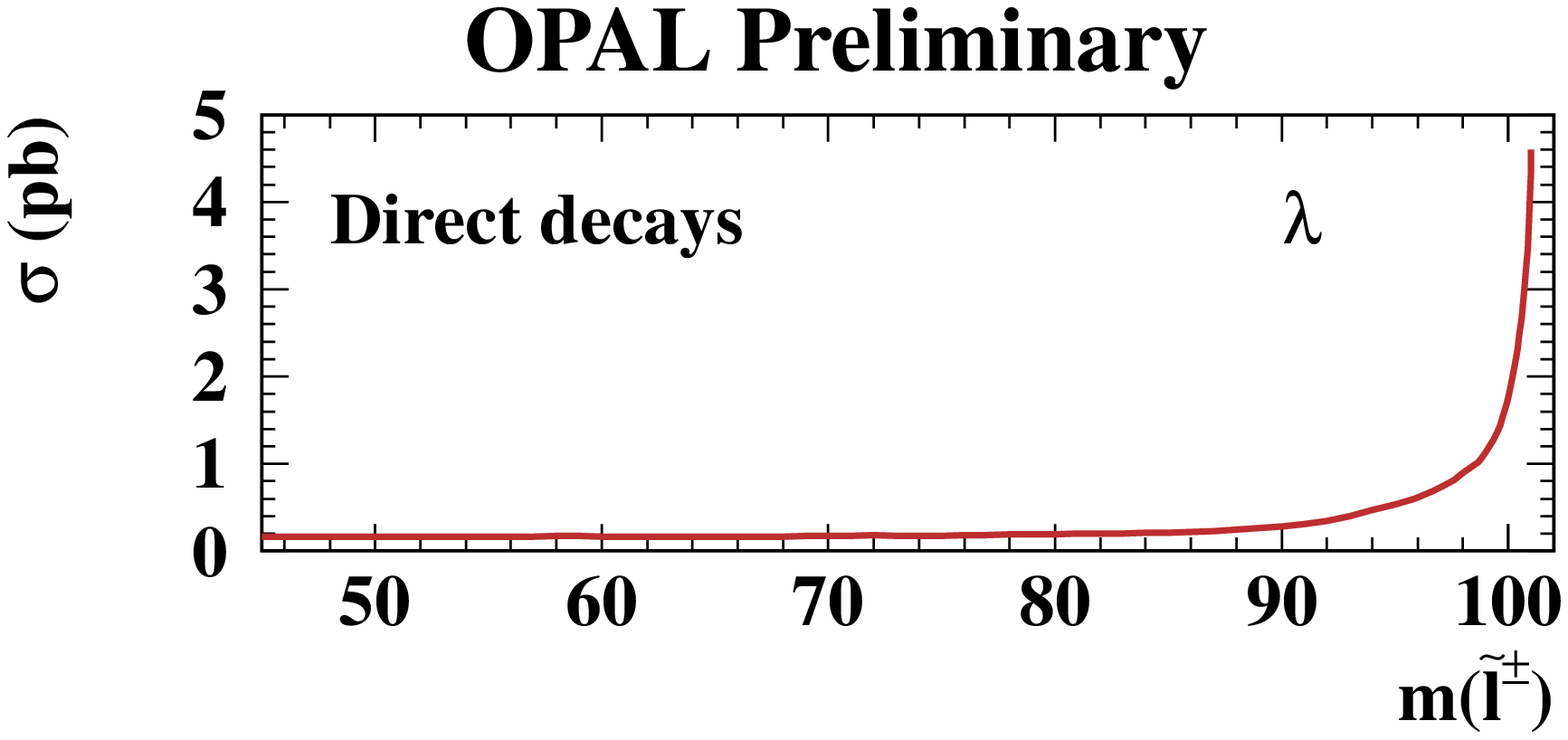} 
\includegraphics*[scale=0.4,clip]{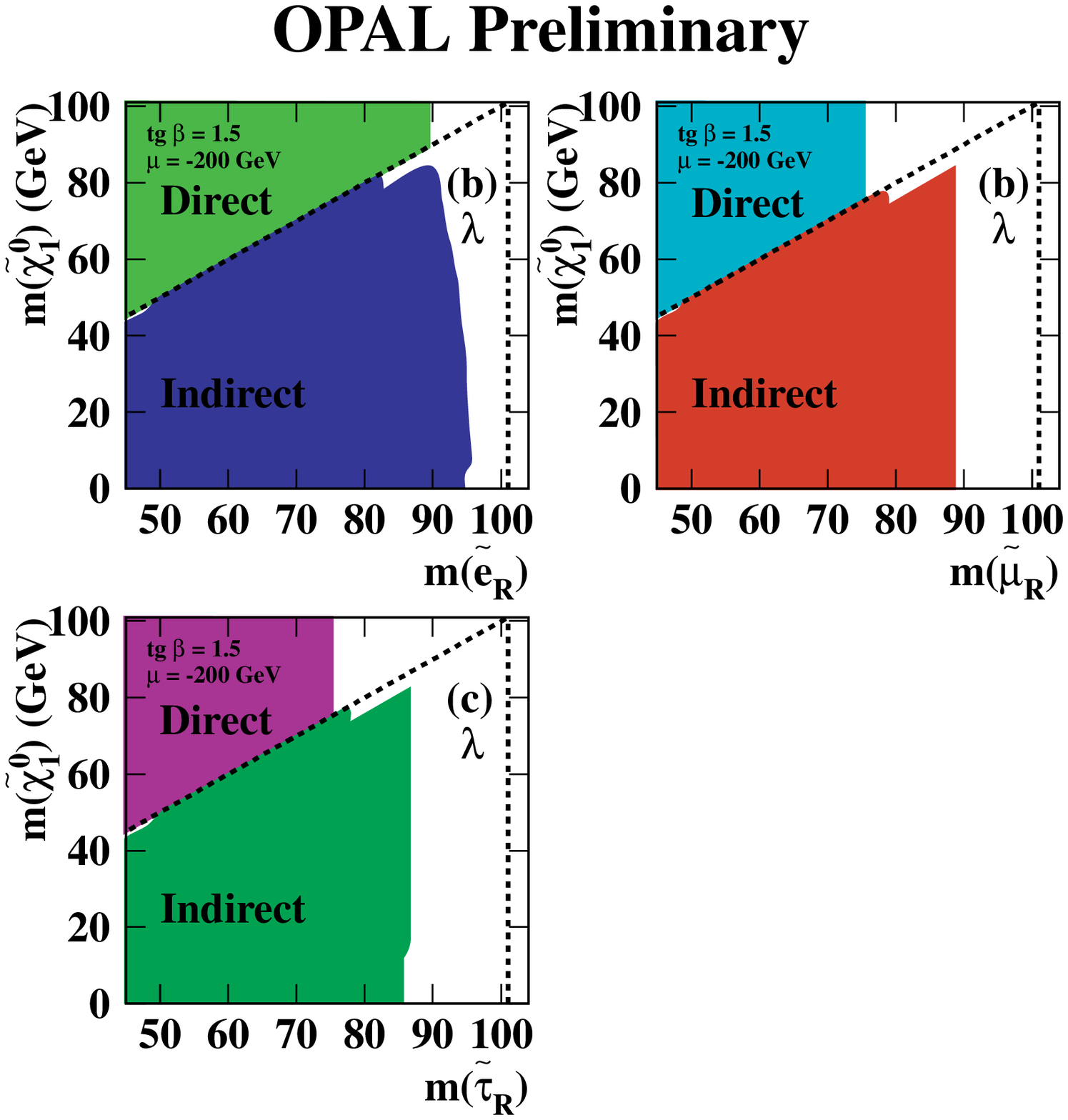} 
\vspace*{-1.7cm}
\end{center}

\hspace*{2.6cm} {\small (a)}
\vspace*{1.25cm}

\fcaption{(a) Upper limit on the
pair-production cross-sections for direct decays. (b) CMSSM 
exclusion regions for selectron pair-production. 
The kinematic limit is shown by dashed lines.}
\vspace*{-10pt}
\end{figure}

The CMSSM interpretation of the results is given in Figure
4(b,c) for smuons and staus.\cite{search2000july} 
For selectrons, due to the higher expected CMSSM
cross-sections, the excluded region is considerably larger.
The mass limits based on the data collected up to 202 GeV are plotted in Figure
5(b).\cite{search1999nov} 

\section{Summary} 
\noindent 
Searches for charginos, neutralinos and scalar
fermions are performed by the OPAL detector. Since no evidence for
supersymmetry with R-parity violation is found, limits are placed on production
cross-sections, sparticle masses and the CMSSM parameter space. 

\nonumsection{References}


\vspace*{-2mm}

\begin{thebibliography}{000}
\bibitem{rpv-sfermion} The OPAL Collaboration, G. Abbiendi {\it et al.},
\epjc{11}{1999}{619}
\bibitem{rpv-gaugino} The OPAL Collaboration, G. Abbiendi {\it et al.},
\epjc{12}{2000}{1}
\bibitem{rpv-lambda} \pn{394}{2 July 1999}{Searches for R-Parity Violating
Decays of Supersymmetric Particles with $\lambda$ Couplings at 189 GeV at LEP}
\bibitem{rpv-lambdap} \pn{411}{12 July 1999}{Searches for R-Parity Violating
Decays of Supersymmetric Particles with $\lambda^\prime$ and
$\lambda^{\prime\prime}$ Couplings at 189 GeV at LEP}
\bibitem{search2000nov} \pn{466}{1 November 2000}{New Particle Searches 
in e$^+$e$^-$ Collisions at $\sqrt{s} = 200 - 209$ GeV}
\bibitem{search2000july} \pn{435}{18 July 2000}{New Particle Searches 
in e$^+$e$^-$ Collisions at $\sqrt{s} = 200 - 209$ GeV}
\bibitem{search1999nov} \pn{418}{5 November 1999}{New Particle Searches 
in e$^+$e$^-$ Collisions at $\sqrt{s} = 192 - 202$ GeV}
\end{thebibliography}
\end{document}